\documentclass[a4paper, amsfonts, amssymb, amsmath, preprint, showkeys, nofootinbib,longbibliography]{revtex4-1}
\usepackage[english]{babel}
\usepackage[utf8]{inputenc}
\usepackage{amsthm}
\usepackage{mathtools}
\usepackage{physics}
\usepackage{xcolor}
\usepackage{graphicx}
\usepackage{siunitx}
\usepackage[left=23mm,right=13mm,top=35mm,columnsep=15pt]{geometry} 
\usepackage{adjustbox}
\usepackage{placeins}
\usepackage[T1]{fontenc}
\usepackage{lipsum}
\usepackage{csquotes}
\usepackage[compat=1.0.0]{tikz-feynman}
\usepackage{feynmp-auto}
\usepackage[version=4]{mhchem}
\usepackage{prettyref}
\usepackage{float}
\usepackage[pdftex, pdftitle={Article}, pdfauthor={Author}]{hyperref} 

\newcommand*{\citesec}[1]{\S~{#1}}
\newcommand*{\citechap}[1]{Ch.~{#1}}
\newcommand*{\citefig}[1]{Fig.~{#1}}
\newcommand*{\citetable}[1]{Table~{#1}}

\newcommand*{\citefootnote}[1]{fn.~{#1}}

\newrefformat{sec}{\citesec{\ref{#1}}}
\newrefformat{fig}{\citefig{\ref{#1}}}
\newrefformat{tbl}{\citetable{\ref{#1}}}
\newrefformat{chap}{\citechap{\ref{#1}}}
\newrefformat{fn}{\citefootnote{\ref{#1}}}
\newrefformat{box}{Box~\ref{#1}}
\newrefformat{ex}{\ref{#1}}

\begin{document}

\title{
MBFormer: A General Transformer-based Learning Paradigm for Many-body Interactions in Real Materials
}
\author{Bowen Hou$^{1,}$}
\email{bowen.hou@yale.edu}
\author{Xian Xu$^{2}$}
\author{Jinyuan Wu$^{1}$}
\author{Diana Y. Qiu$^{1,}$}
\email{diana.qiu@yale.edu}
\affiliation{$^{1}$Department of Materials Science, Yale University, New Haven, CT, 06511, USA}
\affiliation{$^{2}$Department of Applied Physics, Yale University, New Haven, CT, 06511, USA}

\date{\today} 

\begin{abstract}
Recently, radical progress in machine learning (ML) has revolutionized computational materials science, enabling unprecedentedly rapid materials discovery and property prediction, but the quantum many-body problem---which is the key to understanding excited-state properties, ranging from transport to optics---remains challenging due to the complexity of the nonlocal and energy-dependent interactions.
Here, we propose a symmetry-aware, grid-free, transformer-based model, MBFormer, that is designed to learn the entire many-body hierarchy directly from mean-field inputs, exploiting the attention mechanism to accurately capture many-body correlations between mean-field states.  As proof of principle, we demonstrate the capability of MBFormer in predicting results based on the GW plus Bethe Salpeter equation (GW-BSE) formalism, including quasiparticle energies, exciton energies, exciton oscillator strengths, and exciton wavefunction distribution. Our model is trained on a dataset of 721 two-dimensional materials from the C2DB database, achieving state-of-the-art performance with a low prediction mean absolute error (MAE) on the order of 0.1-0.2 eV for state-level quasiparticle and exciton energies across different materials. Moreover, we show explicitly that the attention mechanism plays a crucial role in capturing many-body correlations. Our framework provides an end-to-end platform from ground states to general many-body prediction in real materials, which could serve as a foundation model for computational materials science.

\end{abstract}

\maketitle

\section{Introduction}

The integration of machine learning (ML) and first-principles calculations is reshaping modern computational materials science by dramatically accelerating materials discovery and enabling the study of large, complex systems that were previously computationally prohibitive. Recently, there has been great progress in areas ranging from foundation models for interatomic potentials\cite{deringer2019machine,wood2025family} and learning of ground-state electronic structure\cite{li2024deep, gong2024graph, zhong2023transferable, yu2022time, zhong2022transferable}, to diffusion models generating novel crystal structures\cite{zeni2025generative, gruver2024fine}. Density functional theory (DFT) is by far the most widely used approach for understanding materials' properties from first principles. Recent ML advances, such as graph neural networks (GNN) that can learn DFT tight-binding Hamiltonians\cite{gong2023general,  wang2024universal} with SO(3) equivariance and auto-differentiation (AD) pipelines for electron-phonon coupling in density functional perturbation theory (DFPT) \cite{li2024deep,zhong2023accelerating}, have pushed the frontiers of ML-accelerated calculations for large systems with DFT accuracy. However, because DFT is fundamentally a ground-state, mean-field theory, conventional DFT cannot directly capture properties of the interacting excited-state electrons (the so-called many-body problem) that account for quantitatively accurate band gaps and bandwidths, quasiparticle renormalizations, and electron–hole interactions, which are crucial to understanding materials' optical and electronic properties.

Over the last decades, many-body perturbation theory (MBPT) in condensed matter physics and post-Hartree-Fock methods in quantum chemistry, such as coupled cluster and configuration interaction, have become the gold standard for predicting excited-state properties. In particular, the GW and Bethe Salpeter equation (BSE) formalisms deliver reliable quasiparticle energies and exciton spectra, respectively\cite{onida2002electronic,hybertsen1986electron, Rohlfing2000}. However, these highly accurate quantum many-body methods tend to be extremely computationally expensive. While conventional implementations of DFT scale as $O(N^3)$ with the number of atoms\cite{whitfield2014computational,schuch2009computational}, the GW approximation scales naively as $O(N^4)$\cite{hybertsen1986electron}, the BSE as $O(N^5)$\cite{BGW3,onida2002electronic}, and coupled cluster as $O(N^6)$\cite{bartlett2007coupled}. Meanwhile, ML-based acceleration of excited-state calculations still remains a significant challenge.

Directly transferring the success of DFT Hamiltonian learning schemes to the general many-body problem is challenging because the nearsightedness assumption\cite{prodan2005nearsightedness} and the tight-binding model, on which existing deep Hamiltonian methods\cite{li2022deep} are based, may struggle to capture long-range energy-dependent non-locality in the many-body self energies and interaction kernels. However, a key observation is that in many-body theories, a ground-state calculation often serves as a starting point, and the resulting mean-field wavefunctions are used as building blocks for correlated many-body quantum states, including Green's functions, density matrices, and many-body wavefunctions. Therefore, the entire hierarchy of post-mean-field theories can be thought of as functionals of the set of DFT Kohn-Sham (KS) orbitals, and this functional, could, in principle, be learned through machine learning. Recently, there has been some effort in learning post-DFT quasiparticle bandstructures and spectral functions, which have adopted different approaches for representing the mean field, such as manual optimization of operator features~\cite{knosgaard2022representing,zadoks2024spectral}, GNN for Green's function~\cite{venturella2024unified}, and unsupervised representation learning of DFT KS states~\cite{hou2025data}.
However, these models are by-design limited to a specific type of many-body problem, and there is no unified mechanism to capture the long-range many-body correlations directly from mean-field inputs.

Transformers\cite{vaswani2017attention} have emerged as one of the most powerful architectures in modern machine learning, particularly in natural language processing (NLP), due to attractive features such as parallelism, scalability, and, most notably, the attention mechanism. Here, we develop MBFormer, a novel symmetry-aware transformer-based architecture that can learn the entire many-body hierarchy from ground state mean-field wavefunctions. More specifically, the MBFormer workflow provides a general and customizable end-to-end pipeline from mean-field states to different excited-state tasks, which automatically tokenizes and embeds the high-dimensional KS states, then decodes task-specific tokens to excited-state eigenvalues, eigenstates, or multiparticle interaction kernels within a single, mean-field architecture. For example, a single orbital query yields a quasiparticle energy; an electron–hole pair yields an exciton eigenvector and binding energy; and so forth. 

As a proof-of-principle, we demonstrate the state-of-the-art performance of MBFormer by predicting state-level many-body properties (quasiparticle and exciton properties) within the GW-BSE formalism on a dataset comprising GW-BSE calculations for 721 two-dimensional semiconductors from the C2DB database\cite{gjerding2021recent,haastrup2018computational}. MBFormer demonstrates high accuracy in predicting GW quasiparticle energies across different materials in the test set, achieving a mean absolute error (MAE) of 0.16 eV with a high determinant coefficient of $R^2=0.97$. Moreover, the model enables, for the first time, accurate predictions of two-particle excitonic properties from the single-particle basis, achieving a MAE of 0.20 eV on materials in the test set. The model also exhibits an effective k-grid inference, across exciton energies, oscillator strengths, and exciton wavefunction amplitudes, showing its ability to understand the electron-hole interaction kernel within the BSE.
Lastly, we explore the interpretability of MBFormer and show that the main reason for the success of many-body GW and BSE task is directly derived from the self- and cross-attention mechanisms.
In mapping a general latent representation of mean-field states to arbitrary many-body excitations, MBFormer can serve as a foundation model paradigm for excited-state materials physics.

\section{Results and discussion}\label{band}

\subsection{Theory and Problem Setup}
\label{sec:theory}

\textit{Ab initio} MBPT calculations typically start from a ground state electronic structure obtained from Kohn-Sham DFT. Based on different levels of many-body approximations, the higher-order electron-electron (hole) interaction can be captured by constructing the many-body Hamiltonian $H^{\text{MB}}$
as a matrix in a mean-field basis consisting of $N'$ states selected from the full DFT basis of size $N$.
In general, $H^{\text{MB}} \in \mathbb{C}^{N'\times N'}$ takes the form:

\begin{equation}
    H^{\text{MB}}_{ij} = \langle i | \Hat{H}^{MB}(\{ \phi_{n\textbf{k}}, \epsilon_{n\textbf{k}} \}) | j \rangle,
    \label{eq:MBH}
\end{equation}
where $\{\phi_{n\textbf{k}}, \epsilon_{n\textbf{k}}\}$ is the full set of $N$ DFT Bloch basis states and their corresponding energies. $n$ and $\textbf{k}$ respectively denote the band index and momentum $\mathbf{\textbf{k}}$-point of the Bloch state $\phi_{n\textbf{k}}$. $|i\rangle$ and $|j\rangle$ are the selected $N'$ ground-state basis used to construct the $H^{\text{MB}}$ (with $N' \ll N$), and can represent either single-quasiparticle or multi-particles states such as electron-hole pairs. Then, the excitation energy $E^{\alpha} \in \mathbb{R}^{N'\times 1}$ and excited-state wavefunction $\Phi^{\alpha}(i)=\langle i| \alpha\rangle  \in \mathbb{C}^{N'\times N'}$ can  be obtained by solving the eigenvalue problem: $E^{\alpha}, \Phi^{\alpha}(i) = \text{eigen}(H^{MB}) = \text{eigen}(H^0({E_i})+\Sigma^{\text{MB}}(\{ \phi_{n\textbf{k}}, \epsilon_{n\textbf{k}} \}))$
, where $E_i$ corresponds to eigenvalues of the non-interacting Hamiltonian in the mean-field basis $|i \rangle$, and $\Sigma^{MB}$ is the many-body self-energy operator.
In practice, the DFT KS states are commonly represented by a plane-wave basis ($\phi_{n\textbf{k}}(\mathbf{G}) = \langle \mathbf{G}| n\textbf{k} \rangle \in \mathbb{C}^{d}$, where $d=G_x\times G_y\times G_z$ is the number of plane-waves in three-dimensional space); then, the excited-state eigenvectors and eigenvalues can be expressed as the mapping:

\begin{equation}
    f:\mathbb{C}^{N\times d}, \mathbb{R}^{N\times 1}, \mathbb{C}^{N'\times d^m}, \mathbb{R}^{N'\times 1} \to \mathbb{C}^{N'\times N'}, \mathbb{R}^{N'\times 1}
    \label{eq:mapping}
\end{equation}
,where $m$ denotes the number of particles in the mean-field basis needed to construct the self-energy operator or interaction kernel.
The first two slots of $f$ represents the DFT ground state, while the last two slots of $f$ represents the mean-field starting point of MBPT calculations, which can be single- or multi-particle states and their DFT energies.
$f$ captures the many-body correlation based on different levels of MBPT and returns the interaction-renormalized single- or multi-particle excited states and energies.

\begin{figure*}
    \centering
    \includegraphics[width=\linewidth]{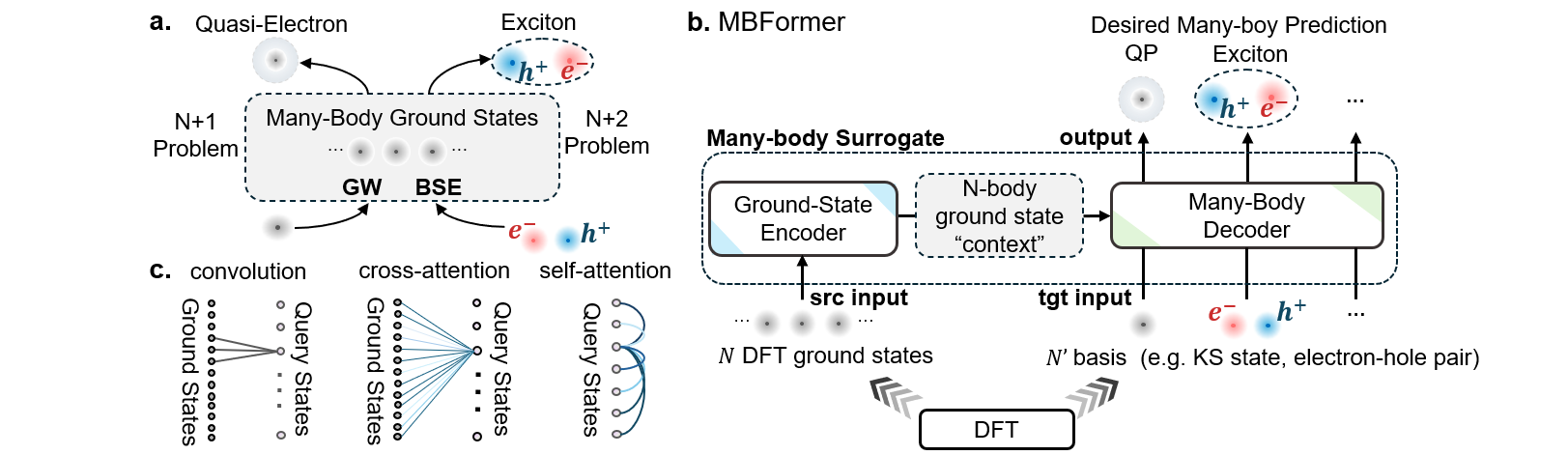}
    \caption{\textbf{Attention mechanism and many-body correlation learning paradigm.} \textbf{a.} Schematic of the formation of excited quasi-particles ($N\pm 1$ problem) and excitons ($N\pm 2$ problem), which can be described by GW and BSE formalisms, respectively. \textbf{b.} MBFormer: an encoder-decoder transformer architecture to learn many-body correlations and excited states. Starting from a DFT calculation, the encoder embeds the mean field information to a latent representation, including correlation among each state guaranteed by the self-attention mechanism. The decoder then predicts excited-state properties, such as energies and wavefunction, by interacting the input target state with the mean field latent representation. \textbf{c.} Schematic of convolution, cross-attention and self-attention mechanisms in predicting and capturing many-body correlation.  }
    \label{fig:scheme}
\end{figure*}

Electron-electron and electron-hole interactions are two key many-body effects in condensed matter physics, giving rise to two fundamental types of elementary excitations: quasiparticles and excitons, respectively. These interactions provide a microscopic understanding of electronic and optical excitations, forming the foundation for a wide range of applications in spectroscopy \cite{Rohlfing2000, brus1984electron, benedict1998optical, Louie1998} and optoelectronics \cite{jailaubekov2013hot, grancini2015role}. Quasiparticles are accurately described by the GW approximation\cite{hedin1965new, hedin1970effects, hybertsen1986electron}, while excitons are captured by the Bethe-Salpeter equation (BSE)\cite{BGW3}. 
Figure~\ref{fig:scheme}(a) illustrates the conceptual formation of a quasiparticle or exciton: the selected independent electron state basis ($|nk\rangle$) or electron-hole pair state ($|cvk\rangle = |ck\rangle \otimes |vk\rangle$) basis interacts with the entire $N$-body ground state environment ($\{\phi_{n\textbf{k}} , \epsilon_{n\textbf{k}}\}$), then forms an excited quasiparticle $|\text{QP}\rangle$ or exciton state $|S\rangle$. However, constructing the MBPT self-energy operator $\Sigma^{\text{MB}}$ is computationally expensive. In practice, the bottleneck is frequently the computation of the interacting polarizability which involves a sum over transition matrix elements between all $N$ states in the mean-field basis\cite{adler1962quantum, wiser1963dielectric}.

Accurately modeling these effects within ML presents several key challenges. First, capturing non-local correlations between the $N'$ basis states and the $N$-body ground state is essential for predicting excited-state properties. Second, the number of states ($N$, $N'$) varies by system, demanding a model that handles variable input and output sizes. Lastly, predictions from DFT require a symmetry-aware, grid-free, and compact representation of high-dimensional KS states.
In NLP, text processing tasks also exhibit input-agnostic characteristics and require ML models to capture dependencies between individual words and the broader context. So, by treating each mean-field basis state as an analog to a “word” or “token” and the ground-state environment as the “context,” the encoder-decoder transformer architecture, along with established NLP training paradigms, can be naturally adapted to address the ML challenges in MBPT described above.
Inspired by this analogy and grounded in MBPT, we design a transformer-based architecture, MBFormer, to learn many-body correlations. As illustrated in Fig.~\ref{fig:scheme}(b), the MBFormer surrogate mirrors the structure of MBPT: starting from a DFT calculation, embedding the ground-state "context", capturing many-body interactions, and ultimately outputing target many-body properties for specified query states.
Fig.~\ref{fig:scheme}(c) shows that attention enables the model to capture global dependencies among all input tokens, while traditional CNN or GNN layers are inherently local. 

\subsection{MBFormer neural network architecture}
\label{sec:mbformer}

\begin{figure*}
    \centering
    \includegraphics[width=\linewidth]{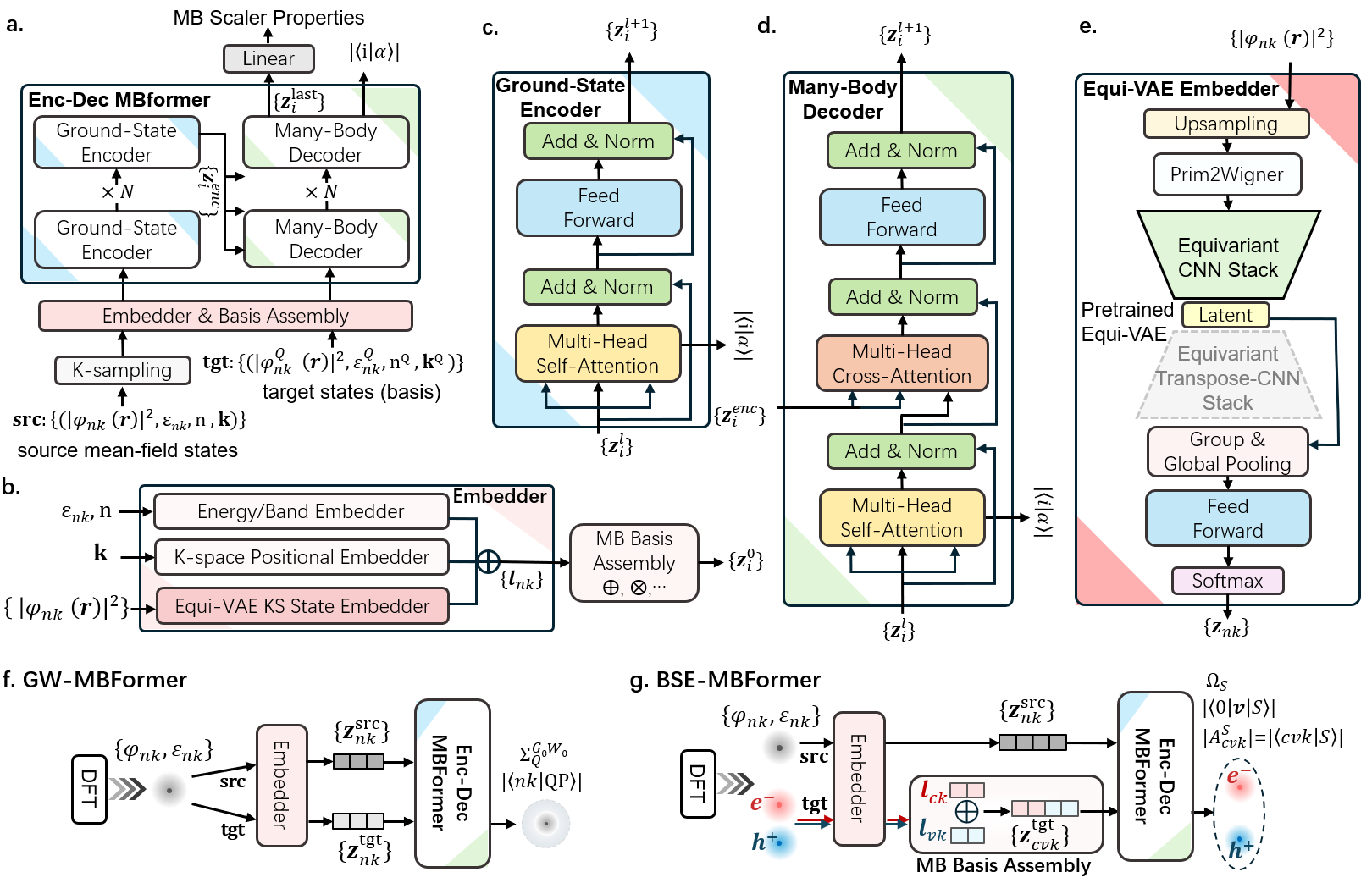}
    \caption{\textbf{Architecture of MBFormer.} \textbf{a.} Overview of the MBFormer architecture, including N-stacked encoder and decoder blocks, which are connected by the cross-attention mechanism. The raw input includes mean-field states information $\textbf{src}=$\{($\phi_{n\textbf{k}}(r)^2$, $\epsilon_{n\textbf{k}}$, $n$, $\mathbf{\textbf{k}}$)\} and target basis state information $\textbf{tgt}=$\{$(\phi^\text{Q}_{n\textbf{k}}(r)^2$, $\epsilon_{n\textbf{k}}^\text{Q}$, $n^\text{Q}$, $\mathbf{\textbf{k}}^\text{Q}$)\}, where $\phi_{n\textbf{k}}(r)$ is the wavefunction, $\epsilon_{n\textbf{k}}$ is the energy, $n$ is the band index, and $\mathbf{\textbf{k}}$ is the crystal momentum (k-point) of mean-field state $|nk\rangle$. The superscript Q means grid-free ``query''.  \textbf{b} The Embedder modules embed information of each individual state stored in \textbf{src} or \textbf{tgt}. The basis assembly module then creates representation of self-energy operator basis ($|nk_1, nk_2,...\rangle$) from the low-dimensional embedded single state information. 
    \textbf{c-d.} The structure of the encoder and decoder blocks. The initial input $\mathbf{z}^{l=0}_i$ is a low-dimensional representation of each individual basis (e.g. independent electron state or electron-hole pair), embedded from the \{($\phi_{n\textbf{k}}(r)^2$, $\epsilon_{n\textbf{k}}$, $n$, $\mathbf{\textbf{k}}$)\} through \textit{Embedder} and \textit{Basis Assembly} modules (see \textbf{b}). $l$ denotes the index of the encoder and decoder layer. The last encoder(decoder) layer output includes standard $\mathbf{z}^{l=\text{final}}_i \in \mathbb{R}^{n\times d_{model}}$ from the \textit{Add} \& \textit{Norm} layer and the attention matrix ($\in \mathbb{R}^{n\times n}$), where $n$ is number of input target basis and $d_{model}$ is dimension of the model (hyper parameter).\textbf{e} The structure of the wavefunction embedder, where E2-VAE integrates E2-CNN stack in its U-net structure\cite{weiler2019general, hou2024unsupervised}. \textbf{f}. Schematic of the workflow of the decoder-encoder MBFormer model for predicting GW energies. \textbf{g} Schematic of the workflow of the decoder-only MBFormer model for predicting excitonic properties.}
    \label{fig:mbformer}
\end{figure*}

Fig.~\ref{fig:mbformer}(a) illustrates the detailed architecture of MBFormer, which is based on an encoder-decoder transformer structure, along with an embedder layer and a basis assembly layer. The model takes two sets of inputs: the $N$ DFT KS states $\textbf{src} = {(\phi_{n\textbf{k}}(r)^2, \epsilon_{n\textbf{k}}, n, \mathbf{\textbf{k}})}$, which are used to embed the many-body ground-state environment, and the selected target KS states $\textbf{tgt} = {(\phi^\text{Q}_{n\textbf{k}}(r)^2, \epsilon^\text{Q}_{n\textbf{k}}, n^\text{Q}, \mathbf{\textbf{k}}^\text{Q})}$, which are used to assemble $N'$ target basis of self-energy operator and predict the desired many-body properties. 
Here, $\phi_{n\textbf{k}}(r)$ denotes the real-space KS wavefunction, $\epsilon_{n\textbf{k}}$ is the KS energy, $n$ is the band index, and $\mathbf{\textbf{k}}$ is the crystal momentum (k-point). 
The model output depends on the specific training task and supports prediction of scalar physical observables ($\mathbb{R}^{N'}$), such as GW self-energy corrections $\Sigma^{GW}_{n\textbf{k}}$, exciton energies $\Omega_S$, and oscillator strength. Additionally, it also supports distribution-like predictions($\mathbb{R}^{N'\times N'}$) , such as exciton wavefunction amplitudes $|A^{S}_{cv\textbf{k}}|$.

First, we preprocess the raw ground-state information from DFT calculations. 
Fig.~\ref{fig:mbformer}(b) shows the embedder and basis assembly modules. The embedder module maps the high-dimensional wavefunction $\phi_{n\textbf{k}}(r)$ into a low-dimensional representation $l_{n\textbf{k}} \in \mathbb{R}^{d_{\text{model}}}$, which also incorporates the corresponding energy- and momentum-information $\{\epsilon_{n\textbf{k}}, n, \mathbf{\textbf{k}}\}$. 
For the KS state embedding, we develop an equivariant variational autoencoder (E2-VAE) architecture (Fig.~\ref{fig:mbformer}(e)) for representation learning of KS states, extending our earlier work that employed a standard VAE for encoding wavefunctions \cite{hou2024unsupervised}. 
The E2-VAE incorporates a U-Net structure augmented with E2-CNN layers \cite{e2cnn}, allowing it to handle coordinate permutations, translations, changes in lattice size, and arbitrary rotations (see more details in SI Note 4).
Next, to encode the position-like metadata $\{\epsilon_{n\textbf{k}}, n, \mathbf{\textbf{k}}\}$ associated with each KS state, we adopt the widely used sinusoidal positional encoding scheme \cite{vaswani2017attention}, which has also been successfully applied to represent reciprocal momentum-space coordinates in prior works \cite{cui2023atomic,gong2024graph} (see SI Note 2).
Lastly, the different self-energy operators $H^{\text{MB}}$ require distinct types of basis, for instance, GW uses single-particle states $|nk\rangle$, while BSE involves electron-hole pairs $|cvk\rangle$. So, an effective representation for multi-particle basis is crucial. Here, taking advantage of the low-dimensionality of $l_{n\textbf{k}}$, the basis assembly module then constructs the self-energy operator basis by combining the low-dimensional representations of individual states, such as through tensor products $z_{i}^0=l_{n_1k_1}\otimes l_{n_2k_2}... \in \mathbb{R}^{d_{\text{model}}\times d_{\text{model}}...}$ or concatenation $z_{i}^0=[l_{n_1k_1}, l_{n_2k_2}, ...] \in \mathbb{R}^{d_{\text{model}}+ d_{\text{model}}...}$.

Fig.~\ref{fig:mbformer}(c-d) illustrates the structure of the encoder and decoder unit blocks, where $\{\mathbf{z}_i^{l}\}$ denotes the hidden representation of the $i$-th output in the $l$-th layer. Here, the attention layer is the core component. Take the self-attention module in $l^{th}$ layer of encoder as an example, the attention output vector ($\in \mathbb{R}^{N_{\text{input}} \times d_{\text{model}}}$) of can be expressed as:

\begin{align}
    \text{Attention}^{l}(\mathbf{z^{l-1}}) &= \text{softmax}\left(\frac{QK^T}{\sqrt{d_\textbf{k}}}\right)V \notag \\
                              &= \text{softmax}\left(\frac{\mathbf{z}^{l-1} \mathbf{W}_Q \mathbf{W}_K^T (\mathbf{z}^{l-1})^T}{\sqrt{d_\textbf{k}}}\right)\mathbf{z}^{l-1} \mathbf{W}_V
    \label{eq:attention}
\end{align}

, where $Q$, $K$, $V \in \mathbb{R}^{N_{\text{input}} \times d_{\text{model}}}$ are the query, key, and value matrices, respectively.
$\mathbf{W}_Q$, $\mathbf{W}_K$, $\mathbf{W}_V \in \mathbb{R}^{d_{model}\times d_{model}}$ are learnable weight matrices that project the input representations $\mathbf{z}^l\in \mathbb{R}^{N_{\text{input}} \times d_{\text{model}}}$ into query, key, and value spaces. The $\text{softmax}(z_i) = \frac{e^{z_i}}{\sum_{j=1}^{N} e^{z_j}}, \quad \text{for } i = 1, \dots, N$ function is applied along the rows, ensuring that the attention matrix weights sums to one for each query. The attention layer computes a weighted sum of the values $V$ based on the similarity between queries $Q$ and keys $K$. This allows the model to selectively focus on relevant parts of all the input data, effectively capturing interactions among the many-body states. The cross-attention mechanism in the decoder works similarly, but it takes the output of the encoder as keys and values, allowing the decoder to attend to the ground-state information while processing the target states.

As a result, the last layer of decoder outputs a scalar vector $f_{\text{FF}}(\text{Attention}^{\text{l=last}})W^{\text{o}} \in \mathbb{R}^{N_{\text{input}} \times 1}$, which thoroughly mixes the input query states and interacts with the ground-state information from the encoder. Here, $f_{\text{FF}}$ is a feed-forward network following attention module and $W^{\text{o}} \in \mathbb{R}^{d_{\text{model}}\times 1}$ is a learnable weight matrix that projects the final decoder output into the desired output space.
More interestingly, the $s^{th}$ row of attention matrix, given by $\text{softmax}(\frac{Q_sK^T}{\sqrt{d_\textbf{k}}}) \in \mathbb{R}^{1\times N_{\text{input}}}$, represents normalized probability distribution over all the key states conditioned on the $s^{th}$ query state, ensured by the softmax operation. This enables the prediction of modulus of the many-body wavefunction, such as exciton or other excited-state wavefunctions, $|\Phi^{\alpha}| \in \mathbb{R}^{1\times N_{\text{input}}}$.

\subsection{Application to one-particle GW prediction}

As an initial application, we evaluate the MBFormer model on predicting quasiparticle energies within the $G_0W_0$ approximation\cite{hybertsen1986electron} (N+1 many-body problem). The quasiparticle energy is given by $E^{\text{QP}}_{n\textbf{k}}=E^{\text{DFT}}_{n\textbf{k}}+\Sigma^{G_0W_0}_{n\textbf{k}}$, where $E^{\text{DFT}}_{n\textbf{k}}$ is KS energy from ground-state DFT, and $\Sigma^{G_0W_0}_{n\textbf{k}}$ is the $G_0W_0$ self-energy correction term. This correction can be expressed:

\begin{equation}
    \Sigma^{G_0W_0}_{n'\textbf{k}'} = \langle n'\textbf{k}'|\Sigma(\{\phi_{n\textbf{k}}, E_{n\textbf{k}}\})|n'\textbf{k}'\rangle
    \label{eq:GW}
\end{equation}

where $\{\phi_{n\textbf{k}}, E_{n\textbf{k}}\}$ denotes the complete set of occupied and unoccupied DFT states, describing all electron scattering channels. Accurately computing this term is computationally demanding, yet essential for converging $G_0W_0$ results\cite{qiu2013optical}. The $|n'\textbf{k}'\rangle$ states are a selected subset of $N'$ target states to which the $G_0W_0$ correction is applied.

Fig.~\ref{fig:mbformer}(f) is an instance of the schematic Fig.~\ref{fig:mbformer}(a), which  illustrates the workflow of the MBFormer for GW prediction (denoted as GW-MBFormer). Starting from a DFT calculation, we partition the DFT KS states into two sets accroding to Eq.~\ref{eq:mapping}: $N$ source states $\{\phi_{n\textbf{k}}, E_{n\textbf{k}}\}$, which are used to construct the ground-state embedding (denoted as src), and $N'$ target query states $\{\phi^\text{Q}_{n'\textbf{k}'}, E^\text{Q}_{n'\textbf{k}'}\}$, which are used to predict the corresponding quasiparticle energies (denoted as tgt). The wavefunction embedder module then maps these high-dimensional source and target states, along with their associated energy-momentum positional information, into compact latent representations $\{\mathbf{z}^{\text{src}(\text{tgt})}_{n\textbf{k}}\}$. 
Then, the output is $N'$ scalar values $\{\Sigma^{G_0W_0}_{n'\textbf{k}'}\}$ corresponding to $N'$ target states. The cross-attention mechanism between MBFormer encoder and decoder allows the target states to interact with the ground-state embedding, enabling the model to capture many-body correlations and accurately predict the $G_0W_0$ self-energy correction term $\Sigma^{G_0W_0}_{n'\textbf{k}'}$.See more training details in SI Note 5.

\begin{figure}
    \centering
    \includegraphics[width=\linewidth]{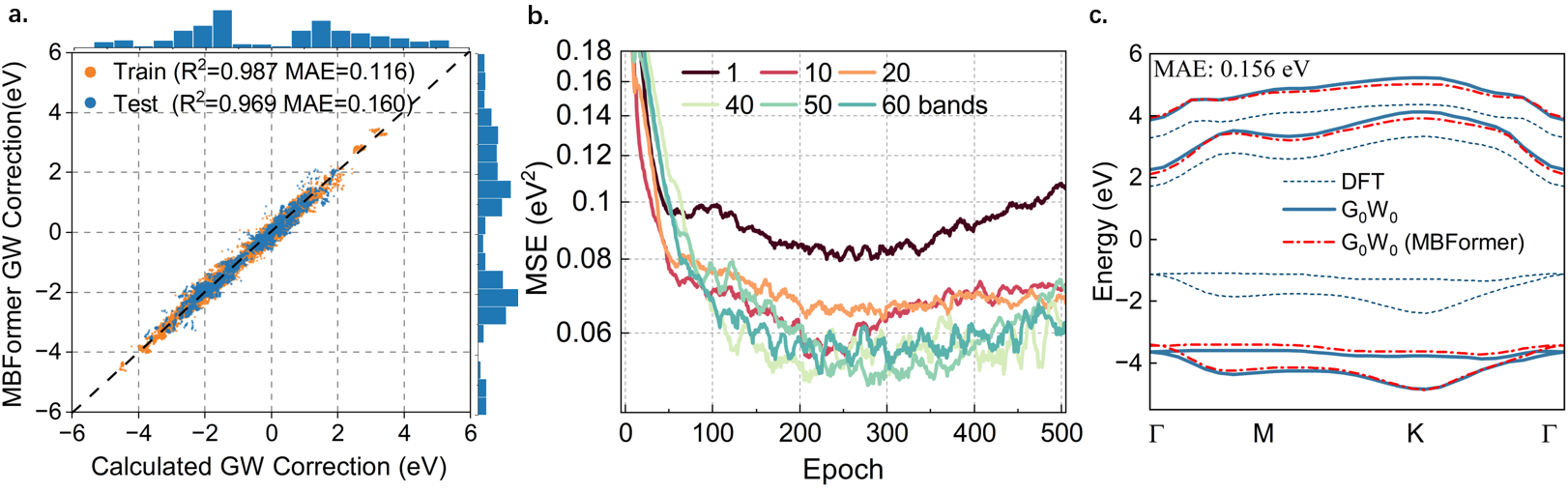}
    \caption{\textbf{MBFormer applicaiton: extrapolation of $G_0W_0$ energies for general 2D materials.} \textbf{a.} Parity plot comparing the exact $G_0W_0$ correction to the machine learning (ML) predicted ones. The histogram reflects the distribution of calculated and ML-predicted GW corrections. The orange(blue) points represent the training(test) dataset. 80\%(20\%) of total 721 2D materials are used for training(test). To better visualize the training trend, a smooth function is applied (see SI for details) \textbf{b.} The validation loss of the MBFormer model during training. The different colored curves represent different number of source states considered in the input. \textbf{c.} The DFT(blue dashed curve), $G_0W_0$ (blue curve) and MBFormer predicted $G_0W_0$ (red dashed line) band structure of monolayer CdSHBr.}
    \label{fig:gw_bands}
\end{figure}

Fig.~\ref{fig:gw_bands}(a) shows the parity plot of the predicted $G_0W_0$ self-energy correction $\Sigma^{G_0W_0}_{n'\textbf{k}'}$ versus the exact values. The GW corrections of the test set have a mean value of 1.36 eV, and their standard deviation is 1.39 eV. The model achieves an MAE of 0.16 eV with a high $R^2=0.969$ across test materials. In addition, within the GW formalism, the convergence of GW quasiparticle energies strongly depends on the number of empty bands \cite{Qiu2013}, which corresponds to the $N$ source states in our ML model. To explore how cross-attention affects this dependency, we train GW-MBFormer with varying numbers of source states. As shown in Fig.~\ref{fig:gw_bands}(b), increasing the number of source states from 1 to 60 conduction bands leads to a substantial reduction in validation loss, with the MSE decreasing from 0.068 to 0.039 (eV$^2$)—a 43\% improvement. 
This suggests the cross-attention mechanism plays a key role in capturing essential many-body correlations between source and target states. This observed behavior also aligns with physical intuition: including more mean-field correlations enhances the prediction of quasiparticle properties, resulting in faster convergence during the training of MBFormer. Lastly, Fig.~\ref{fig:gw_bands}(c) presents the MBFormer-predicted GW band structure for monolayer CdSHBr from the test set, showing excellent agreement with the reference $G_0W_0$ calculation (see more band structure predictions in SI Note 5). Notably, MBFormer inherently produces a smooth band structure, in contrast to previous GW ML models \cite{knosgaard2022representing} that do not incorporate generative modeling.

\subsection{Application to two-particles BSE prediction: universal materials inference}

Next, we move to excitonic property prediction, which can be more complicated due to the much larger Hilbert space. Exciton eigenstates and eigenvalues are obtained by diagonalizing an effective two-particle Hamiltonian including both the mean-field band energy term and BSE electron-hole interaction kernel $K$. The BSE Hamiltonian takes the form\cite{BGW1, Rohlfing2000}:
\begin{equation}
H^{\text{BSE}}_{cvk, c'v'\textbf{k}} = \langle vck|E_{c\textbf{k}}-E_{v\textbf{k}}|vck \rangle \delta_{cc'}\delta_{vv'} + \langle vck|K|v'c'k \rangle
\label{eq:BSE_Hamiltonian}
\end{equation}
where $|vck\rangle=|vk\rangle \otimes |ck\rangle$ is an electron-hole pair, and $E_{n\textbf{k}}$ is the energy of the noninteracting Bloch state $|nk\rangle$, which can be taken from either DFT or GW. 
Accurate ML models for optical property prediction are essential for high-throughput materials screening, yet remain a significant challenge. Recently, the GNNopt model \cite{hung2024universal} demonstrated the ability to directly predict optical absorption spectra, which however is limited to a fixed energy window and a discretized grid. Also, its model architecture that directly predicts the absorption spectra offers no explicit prediction of the individual excitonic energy levels, which are essential for accurately characterizing underlying optical properties.
The MBFormer model, designed to directly learn the many-body correlations inherent in both the electron-hole Kernel and the screened Coulomb interaction can easily be extended to electron-hole pairs, in a model we refer to as BSE-MBFormer, and thus, predict excitonic properties for individual exciton state.

Fig.~\ref{fig:mbformer}(g) illustrates the BSE-MBFormer model for predicting excitonic properties. Starting from a DFT calculation, the relevant electron and hole states near the Fermi level, which primarily contribute to the bound exciton within the band gap, are first embedded into a low-dimensional representation $l_{c\textbf{k}}$ and $l_{v\textbf{k}}$. The MB Basis Assembly module then constructs a representation of the electron-hole pair, $z_{cv\textbf{k}}$, by combining these low-dimensional single particle embeddings through a simple concatenation operation: $z_{cv\textbf{k}}=l_{c\textbf{k}}\oplus l_{v\textbf{k}}$. We refer to $z_{cv\textbf{k}}$ as the target electron-hole pair (tgt in Fig.~\ref{fig:mbformer}(g)), as they represent the electron-hole pairs that are used to build the BSE Hamiltonian and calculate desired excitonic properties. Similar to GW-MBFormer, a portion of DFT calculated KS states  $\{\phi_{n\textbf{k}}, E_{n\textbf{k}}\}$ are also used to construct the mean-field embedding, which is denoted as the source states (src in Fig.~\ref{fig:mbformer}(g)), and the cross-attention mechanism allows the target electron-hole pairs to interact with the mean-field embedding.

\begin{figure}
    \centering
    \includegraphics[width=\linewidth]{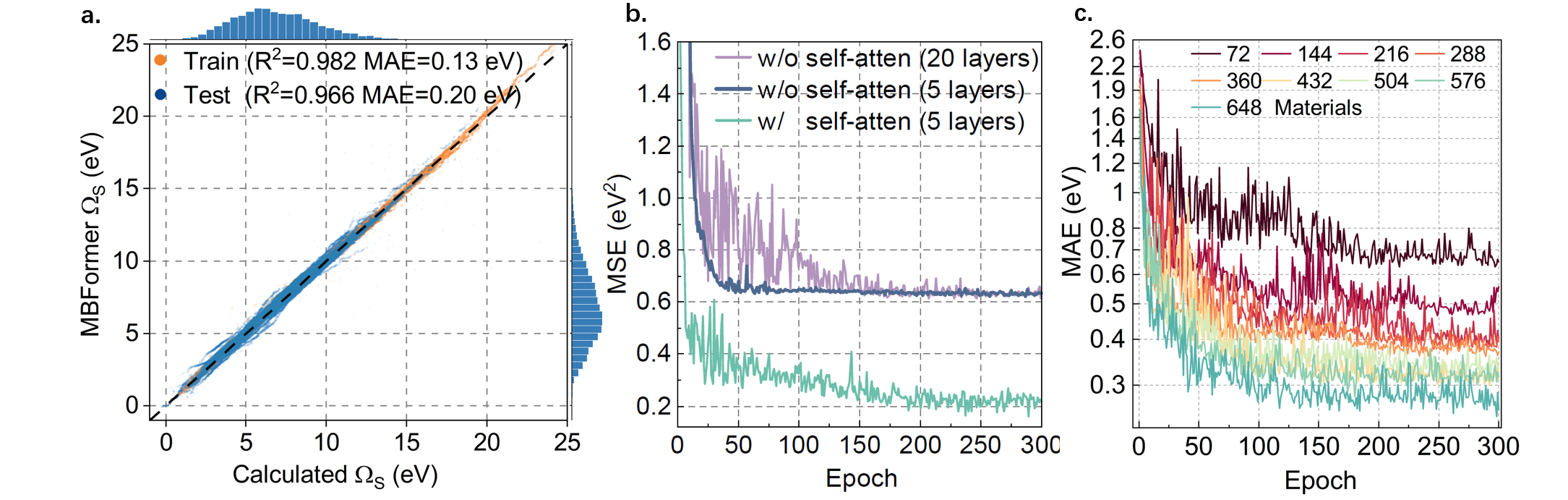}
    \caption{\textbf{MBFormer applicaiton: extrapolation of excitonic properties for general 2D materials.} \textbf{a.} Parity plot comparing the exact calculated exciton energies to the machine learning (ML) predicted exciton energies. The orange(blue) points represent the training(test) dataset. 90\%(10\%) of total 721 2D materials are used for training(test). \textbf{b.} The validation MSE of BSE-MBFormer during training. The blue and green curves denote 5-layer BSE-MBFormer training without and with the self-attention module. The purple curve denotes 20-layers BSE-MBFormer without self-attention module. \textbf{c.} The validation MAE of the BSE-MBFormer during training. The different colored curves represent different numbers of materials in the training set, ranging from 72 to 648. }
    \label{fig:bse_extrapolation}
\end{figure}

As a proof of concept, we start by evaluating the extrapolation capability of BSE-MBFormer for predicting the energy of each exciton state for a diverse set of 2D materials. Fig.~\ref{fig:bse_extrapolation}(a) shows a parity plot of the predicted exciton energies from BSE-MBFormer, where the MAE of 72 materials from a randomly selected test set (10\% of total dataset) is 0.20 eV with $R^2=0.966$. For the first 100 low-energy excitons for each materials, the MAE is reduced to 0.15 eV, suggesting that the model is more accurate for bound exciton states. 
Fig.~\ref{fig:bse_extrapolation}(b) presents the validation MSE during training, comparing models with and without the self-attention mechanism. Remarkably, incorporating self-attention reduces the MSE from 0.63 to 0.16 eV$^2$ (a 75\% improvement) and lowers the corresponding MAE from 0.56 eV to 0.21 eV (a 62.5\% improvement). Additionally, the purple curve illustrates that simply increasing the number of MBFormer layers, without introducing self-attention, eventually converges to the performance of a 5-layer model lacking attention. This result provides strong evidence that architectural improvements, particularly the incorporation of correlations through self-attention, play a crucial role in enhancing many-body prediction accuracy, while simply increasing the depth of model has little effect. Fig.~\ref{fig:bse_extrapolation}(c) presents the training dynamics of BSE-MBFormer with different sizes of training datasets, which clearly shows that the model hasn't reached to its architectural limit and can learn more underlying correlations by increasing training materials. We note that the BSE-MBFormer results presented here only contain self attention. We also develop a model paralleling GW-MBFormer with both self- and cross-attention (see SI Section Note 1). However, due to the delocalized nature of exciton wavefunctions compared to the size of the unit cell, they are less sensitive to short-range fluctuations that are captured by the contribution of higher-energy states to the polarizability~\cite{qiu2016screening}. Hence, self-attention within the basis used the construct the BSE Hamiltonian is sufficient to model the electron-hole interaction.

\subsection{Application to two-particles BSE prediction: fine k-grid inference}

\begin{figure}[H]
    \centering
    \includegraphics[width=\linewidth]{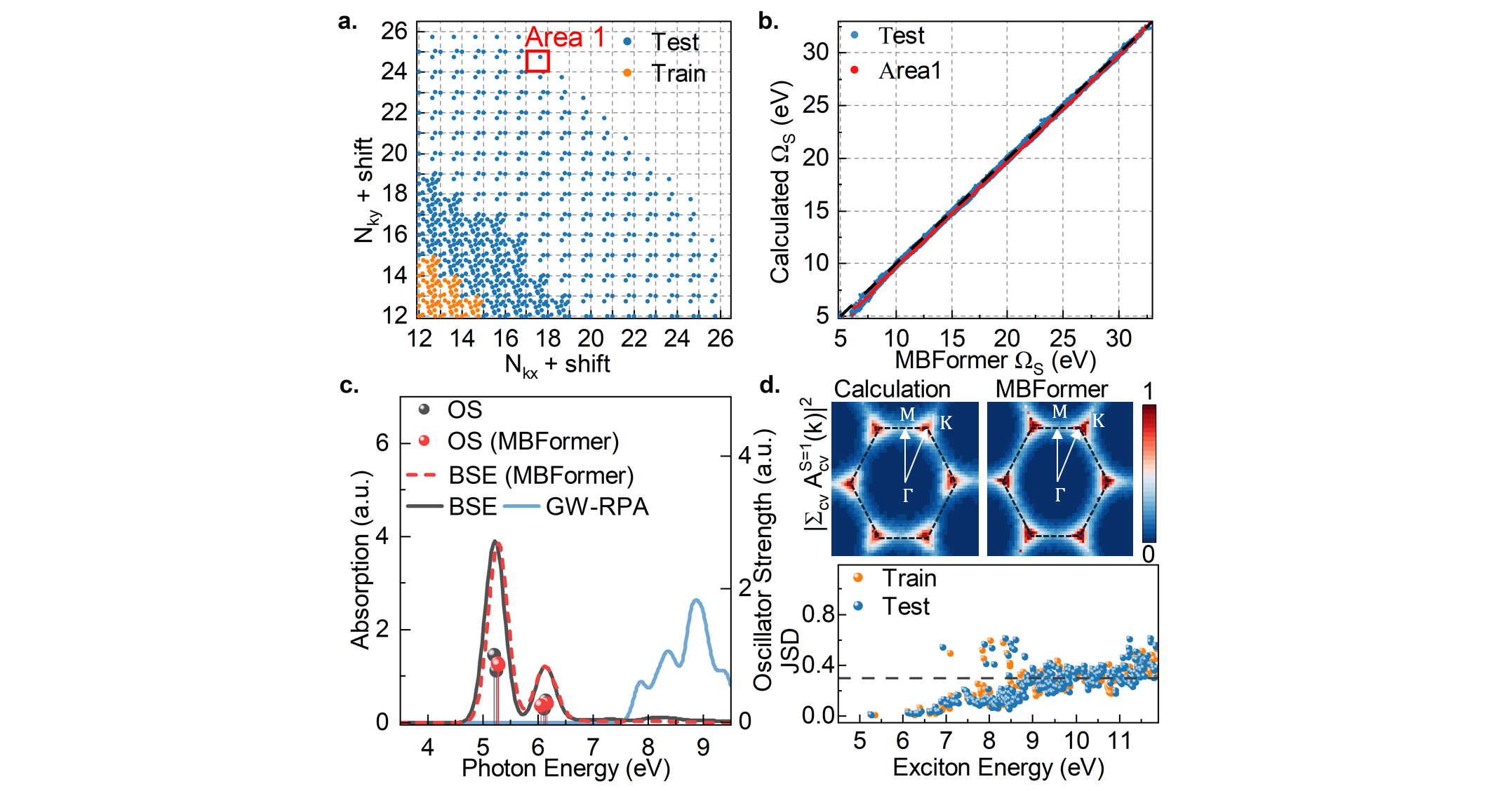}
    \caption{\textbf{Inference of excitonic properties from coarse to fine k-grid sampling.} \textbf{a.} An overview of calculated augmented dataset (900 datapoints) of monolayer hBN used to train the MBFormer model. Each individual point stands for a BSE Hamiltonian constructed from a specific k-grid sampled over the 1st Brillouin zone. The integer part of the x(y)-axis represents the number of k-points sampled along $k_x(k_y)$-direction in reciprocal space. The fractional part of coordinate represents a small shift applied to the k-grid. The orange (blue) dots represent the training (test) dataset. \textbf{b} parity plot comparison of exciton energies for calculated and BSE-MBFormer predicted exciton energies. The blue dots denote comparison for all test sets and red dots denote comparison for area 1 in \textbf{a}.
    \textbf{c.} The BSE calculated absorption spectrum (black solid curve) and the MBFormer predicted absorption spectrum (red dashed curve) of monolayer hBN in area 1. The black (red) dots represent the exciton energies and corresponding oscillator strengths calculated from BSE (predicted by MBFormer). \textbf{d.} The BSE calculated wavefunction amplitudes of the first exciton in hBN 
    (left) and the MBFormer predicted exciton wavefunction amplitudes 
    (right) of monolayer hBN in area 1. The lower panel shows the Jensen–Shannon divergence (JSD) of the ML predicted exciton wavefunction amplitudes of the training (orange) and test (blue) dataset.  }
    \label{fig:bse_interpolation}
\end{figure}

Next, to further explore the ability of the attention mechanism, we evaluate the inference capability of BSE-MBFormer for fine k-grids, using a model trained from coarse k-grids. This is a key point because BSE calculations converge very slowly with respect to k-point sampling, making the generation of converged datasets for training a significant bottleneck. We focus on monolayer hexagonal boron nitride (hBN) as a test case and explore three types of BSE excitonic predictions: scaler value prediction of exciton energies $\Omega_S$ and exciton oscillator strength $|\langle 0|v|S \rangle|$; and distributional vector prediction of the modulus of the exciton wavefunction $|A^{S}_{cv}(k)|$.

Fig.~\ref{fig:bse_interpolation}(a) illustrates the data augmentation strategy for BSE (see more details in SI Note 4). The total dataset consists of 900 different BSE Hamiltonians of monolayer hBN sampled from different k-grid, where the integer part of each point's coordinate stands for the number of k-points sampled along that direction in 1BZ, while the fractional part represents a small shift applied to the k-grid. Here, we train the model on 6 coarse k-grids with 84 data augmentations (10\% of the dataset), represented by the orange dots in Fig.~\ref{fig:bse_interpolation}(a). 
As shown in Fig.~\ref{fig:bse_interpolation}(b), the blue parity plot compares the exact exciton energies with those predicted by BSE-MBFormer on the test set, demonstrating good agreement with a low MAE of 0.15 eV and a high $R^2$ value of 0.999. Additionally, we highlight region 1 for further comparison, as it lies well outside the coarse k-grid regime, which also achieves a high $R^2=0.998$.
This suggests that the self-attention mechanism can effectively interpolate the unseen electron-hole pair basis and learn the underlying correlation among them.
Additionally, the positional embedding scheme (see SI Note 2) plays a crucial role by enabling the model to capture the meta-information of {$\textbf{z}_{cv\textbf{k}}$} in both energy and momentum space, particularly in the absence of a predefined ordering of the input.

Similar to exciton energy prediction, MBFormer can also be trained to predict the dipole oscillator strength of each exciton, denoted as $|\langle 0|v|S \rangle|$ ($\mathbb{R}^{N'\times 1}$). 
However, this task is challenging due to the degeneracy of exciton states, which introduces a random rotation over the degenerate subspace, which may be particularly prominent for higher-energy excitons where there is a dense continuum of nearly degenerate states. 
For training, we first apply max-renormalization to scale the dipole strength of each exciton to a maximum of 1. The training achieves a low MAE of 0.003 on the test set. By combining the predicted dipole strengths and exciton energies, we can reconstruct the excitonic absorption spectrum as $\epsilon_2(\omega)\propto \sum_S|\textbf{e}\cdot\langle 0|v|S \rangle|^2\delta(\omega-\Omega_S)$\cite{BGW1}. Fig. \ref{fig:bse_interpolation}(c) presents the predicted absorption spectrum from area 1 with a dipole MAE of 0.003, where the predicted spectrum (red dashed line) closely matches the one computed via the BSE (black solid line). The blue curve shows the spectrum without electron-hole interaction.

Lastly, we address a more challenging prediction of distributional data, specifically, the exciton wavefunction amplitudes $|A^{S}_{cv\textbf{k}}|$, meaning MBFormer must learn a mapping from a low-dimensional latent space ($d$) to a high-dimensional distribution over $N'$ states, i.e., $f: \mathbb{R}^{N'\times d} \to \mathbb{R}^{N'\times N'}$  ($N' \gg d$).  Given the distributional nature of $|A^{S}_{cv\textbf{k}}|$, we adopt the Jensen-Shannon divergence (JSD) as our loss function (see SI). The first row of Fig.~\ref{fig:bse_interpolation}(d) compares the ground truth and predicted $|A^{S}_{cv\textbf{k}}|$ of the first exciton from area 1. The k-indices in $|A^{S}_{cv\textbf{k}}|$ are reordered to match the Wigner cell of 1BZ and the two distributions match well with each other. 
This highlights that, aided by the KS state embedding and the positional encoding scheme, the global attention mechanism of MBFormer effectively captures long-range correlation patterns among the input states. The second row shows the JSD values of the predicted exciton wavefunction amplitudes across training (orange, 12$\times$12$\times$1) and test (blue, area 1) datapoint. As expected, lower band-edge excitons yield low JSD (below 0.05), while higher-energy excitons exhibit larger JSD due to increased electron-hole mixing and degeneracy randomness. In conclusion, the k-grid inference ability for high-dimensional data such as wavefunctions establishes a foundation for the self-energy operator learning in future work.

\section{Conclusion}\label{sec:conclusion}

In summary, we present MBFormer, a symmetry-aware, transformer-based end-to-end pipeline designed for general many-body predictions in real materials. By partitioning the DFT KS states into source and target sets, our novel training paradigm consciously mimics the workflow of post-DFT many-body approaches. This approach effectively addresses three core challenges in the prediction of many-body physical observables: (i) introducing and validating attention mechanisms specifically designed to capture complex many-body correlations, (ii) enabling a grid-free workflow with input-output flexibility regarding the number of basis states used in self-energy operators, and (iii) learning unsupervised, symmetry-invariant representations of single- and multi-particle ground state basis. Our model is, in principle, adaptable to any many-body interactions, but as an initial proof of principle, we demonstrate its application to the N+1 particle problem, through predicting GW corrections, and the N+2 particle problem for predicting exciton properties through the BSE formalism. Additionally,  we demonstrate the inference capability of MBFormer from coarse to fine k-grid sampling for BSE, including scalar prediction such as exciton energies and oscillator strengths, and distribution predictions of exciton wavefunction amplitudes. Our results directly demonstrate that introduction of an attention mechanism is crucial for many-body tasks, significantly increasing the prediction accuracy of both GW and BSE.
We hope this work encourages further exploration of machine learning applications in many-body physics and paves the way for developing foundation models in materials science.

\section{Acknowledgement}
This work was primarily supported by the U.S. Department of Energy, Office of Science, Basic Energy Sciences under Early Career Award No. DE-SC0021965. Development of the BerkeleyGW code was supported by Center for Computational Study of Excited-State Phenomena in Energy Materials (C2SEPEM) at the Lawrence Berkeley National Laboratory, funded by the U.S. Department of Energy, Office of Science, Basic Energy Sciences, Materials Sciences and Engineering Division, under Contract No. DE-C02-05CH11231. Calculations on benzene and black phosphorus were supported by the National Science Foundation Division of Chemistry under award number CHE-2412412. The calculations used resources of the National Energy Research Scientific Computing (NERSC), a DOE Office of Science User Facility operated under contract no. DE-AC02-05CH11231; the Advanced
Cyberinfrastructure Coordination Ecosystem: Services \& Support (ACCESS), which is
supported by National Science Foundation grant number ACI-1548562; and the Texas Advanced Computing Center (TACC) at The University of Texas at Austin.

\bibliography{materials}

\end{document}